\documentclass[twoside,american]{elsarticle}
\usepackage[T1]{fontenc}
\usepackage[latin9]{inputenc}
\usepackage{float}
\usepackage{url}
\usepackage{amsmath}
\usepackage{amsthm}
\usepackage{amssymb}
\usepackage{graphicx}

\makeatletter

\floatstyle{ruled}
\newfloat{algorithm}{tbp}{loa}
\providecommand{\algorithmname}{Algorithm}
\floatname{algorithm}{\protect\algorithmname}

\theoremstyle{plain}
\newtheorem{thm}{\protect\theoremname}
\theoremstyle{definition}
\newtheorem{defn}[thm]{\protect\definitionname}
\theoremstyle{remark}
\newtheorem{claim}[thm]{\protect\claimname}
\theoremstyle{plain}
\newtheorem{prop}[thm]{\protect\propositionname}
\theoremstyle{definition}
\newtheorem{example}[thm]{\protect\examplename}

\journal{arxiv.org}

\usepackage{algorithm}
\usepackage{caption}

\usepackage{algorithmicx}
\usepackage{algpseudocode}
\usepackage{varwidth}

\newcommand{\alglinenoNew}[1]{\newcounter{ALG@line@#1}}

\newcommand{\alglinenoPush}[1]{\setcounter{ALG@line@#1}{\value{ALG@line}}}

\usepackage{babel}
\providecommand{\definitionname}{Definition}
\providecommand{\theoremname}{Theorem}



\makeatother

\usepackage{babel}
\providecommand{\claimname}{Claim}
\providecommand{\definitionname}{Definition}
\providecommand{\examplename}{Example}
\providecommand{\propositionname}{Proposition}
\providecommand{\theoremname}{Theorem}

\begin{document}
\begin{frontmatter}
\title{My part is bigger than yours - assessment within a group of peers\tnoteref{t1,t2}}
\author[waiibkis]{Konrad Ku\l akowski\corref{cor1}}
\ead{konrad.kulakowski@agh.edu.pl}
\author[wms]{Jacek Szybowski}
\ead{jacek.szybowski@agh.edu.pl}
\cortext[cor1]{Corresponding author}
\address[waiibkis]{AGH University of Krakow, WEAIiIB, Applied Computer Science Department}
\address[wms]{AGH University of Krakow, Faculty of Applied Mathematics}
\begin{abstract}
A project (e.g., writing a collaborative research paper) is often
a group effort. At the end, each contributor identifies their contribution,
often verbally. The reward, however, is very frequently financial.
It leads to the question of what (percentage) share in the creation
of the paper is due to individual authors. Different authors may have
various opinions on the matter; even worse, their opinions may have
different relevance. In this paper, we present simple models that
allow aggregation of experts\textquoteright{} views, linking the priority
of his preference directly to the assessment made by other experts.
In this approach, the more significant the contribution of a given
expert, the greater the importance of his opinion. The presented method
can be considered an attempt to find consensus among peers involved
in the same project. Hence, its applications may go beyond the proposed
study example of writing a scientific paper. 
\end{abstract}
\end{frontmatter}

\section{Introduction}

From time to time, we observe behind-the-scenes discussions about
which author in a scientific paper is the most relevant. The answers
can vary. First, last, correspondence, first three authors, it doesn't
matter, etc. At its core, it is a question about the value of the
contribution and the effort that an author has made to the paper.
Scientific journals help answer this question by making it possible
to determine the workload of a given author by describing descriptively
what that author did. Such a description makes it possible, to some
extent, to form an opinion about the significance of an author's contribution.

However, there are times when such a description is insufficient,
and a precise percentage determination of \textquotedbl contribution
to the work\textquotedbl{} for each author separately is necessary.
This is the case when a certain amount of money is awarded for creating
a work to be distributed among the authors. In such a situation, the
\textquotedbl distribution of percentages\textquotedbl{} is often
done dictatorial, with the corresponding author playing the role of
a dictator. In a situation where the authority of all authors is comparable,
or worse, where the corresponding author is a less recognized researcher
than some other researchers on the author list, misunderstandings
and irritations can quickly arise. On the one hand, the correspondent
(or lead) author should have the best knowledge of the other authors'
contributions to the work. On the other hand, other authors of greater
esteem can easily question their decisions.

Adopting arbitrary regulations to distribute the award may be a way
out of this somewhat uncomfortable situation. For example, the prize
can be divided equally among all participants, or rigid proportions
can be set for the award for the first and subsequent authors. Unless
they arouse personal animosity between the authors, these solutions
leave a feeling of injustice. The question, therefore, arises as to
whether a dictatorial approach, on the one hand, and establishing
a rigid framework for the distribution of the prize, on the other,
can be avoided. With a team of peers, this is possible.

In this paper, we propose an approach based on aggregating the opinions
of individual authors/experts using prioritization of views. Since
team members who were more involved in the project usually have a
better understanding of the project and the contribution of other
team members to the work done, prioritization should be related to
the size of this contribution. In this way, we avoid a situation where
those whose participation is small or marginal have the same share
in the decision as those whose participation is significant. Of their
significant and large contribution, the latter are given greater priority
in determining the award distribution.

In doing so, we assume that the work took place in a team of peers,
i.e., that there is no obvious dictator whose opinion everyone naturally
aligns with and that there is a real opportunity for each team member's
contribution to be assessed by others. The absence of a dictator results
in the need for team decisions on the prize distribution. The possibility
of mutual evaluation, on the other hand, enables group decision-making.
It is the case, for example, when a team of researchers of similar
experience and reputation works together on a project, having stand-up
meetings regularly where the work progress is discussed.

The problem outlined above applies to more than just creating scientific
articles. A similar situation occurs with the distribution of awards
for various other team achievements, such as grants, organizational
projects, or software projects in IT companies. The solution proposed
in this paper is based on the aggregation procedure used in the group
decision-making (GDM) process, which is used in the pairwise comparison
method. The basics of the classical aggregations method are described
by Saaty and Aczel \citep{Aczel1983pfsr}, Saaty \citep{Saaty1989gdma}
and later by Forman and Peniwati \citep{Forman1998aija}. The problem
of aggregating expert opinion for the pairwise comparison method has
been studied many times. Aggregation procedures have been adapted
for models using fuzzy sets \citep{Coffey2021idog}, hesitant probabilistic
fuzzy linguistic sets \citep{Singh2020ahpf} or intervals \citep{Zadnik2013apig}.
Among GDM studies, consensus-building methods have a special place.
In \citep{Lin2022aotn} a method is proposed that considers the inconsistency
of expert judgments and acceptable consensus in parallel. Gong et
al. \citep{Gong2024mfac} consider the problem of consensus in social
networks, where an additional element that can help or hinder consensus
building is the moderator's actions. The proposed maximum fairness
consensus model uses decision-maker fairness assessment inspired by
social comparison and the Gini coefficient. Large-scale group decision-making
(LSGDM) is the subject of another paper \citep{Liang2024acmc}. Liang
et al. propose the framework for LSGDM, including clustering, discovering
and managing manipulative behavior, and consensus reaching. Zhao et
al. \citep{Zhao2023cmwi} pay attention to individual participation
and satisfaction in decision-making. They propose a new consensus
model based on the optimization of utility aggregated using additive
Choquet integral. The consensus issue, often understood as the minimization
of the difference between decision-makers preferences and the collective
opinion, is described in more detail in \citep{Dong2016cbig}.

The pairwise comparison of alternatives approach has also resulted
in several studies in group decision-making. Traditionally, there
are two opinion aggregation procedures in the Analytic Hierarchy Process
(AHP) \citep{Saaty1977asmf} method: aggregation of individual judgments
(AIJ) and aggregation of individual priorities (AIP) \citep{Aczel1983pfsr,Saaty1998rbev,Forman1998aija}.
We will present the latter in more detail in Section \ref{subsec:Multiplicative-profile-aggregati}.
Both procedures are commonly accepted and used for preference aggregation
given in the form of pairwise comparisons (PC) matrices (see Sec.
\ref{subsec:Pairwise-comparisons}), whereby AIJ is recommended when
experts have previously agreed on a common assessment framework \citep{Forman1998aija}.
Ramanathan and Ganesh \citep{Ramanathan1994gpam} argue in favor of
AIP as it meets the Pareto principle when using the eigenvector prioritization
method (EVM) \citep{Saaty1977asmf}, whereas AIJ does not. Both methods
are equivalent when using the geometric mean method (GMM) (or logarithmic
least square method (LLSM)\footnote{Both GMM and LLSM are equivalent for complete and incomplete PC matrices
\citep{Kulakowski2020utahp}.} for the ranking calculation \citep{Barzilai1994arrn}. In addition
to the two mentioned, other methods of aggregating preferences have
emerged. The group Euclidean distance (GED) minimization method has
been proposed by Blagojevic et al. \citep{Blagojevic2016haoi}. The
process of progressively modifying individual experts' preference
matrix to minimize the defined geometric consensual consistency index
(GCCI) lies at the heart of the idea from Dong et al. \citep{Dong2010cmfa}.
Altuzarra et al. proposed Bayesian prioritization (BPP), allowing
the aggregation procedure to be extended to preferences (utilities,
priorities) between the alternatives that are given as probability
distributions \citep{Altuzarra2007abpp}. Escobar and Moreno-Jimenez
consider \citep{Escobar2007aoip} a new aggregation method called
Aggregation of Individual Preference Structures (AIPS). Aggregating
more diverse preferential information allows phenomena such as judgment
uncertainty, problem vision, or interdependence to be considered.
Lin and Kou \citep{Lin2015brot} focus on the Bayesian revision method,
which allows for improving PC matrices. Lin et al. \citep{Lin2022aotn}
propose aggregation of the nearest consistent matrices with the additional
adjustment connected with the acceptable consensus. Kulakowski et
al. \citep{Kulakowski2024rhao} introduce an aggregation method equipped
with heuristically adjusted priorities, preventing manipulative behavior
by some decision-makers.

\section{Preliminaries\protect}\label{sec:Preliminaries}

\subsection{Pairwise comparisons\protect}\label{subsec:Pairwise-comparisons}

Many methods including AHP, HRE, MACBETH, and BWM, outranking methods
like, e.g., PROMETHEE \citep{Kulakowski2020utahp,Greco2016mcda} and
various hybrid solutions use pairwise comparisons (PC) as a source
of information about decision makers preferences. The approach has
its extensions to cover different data representations including intervals
\citep{Grosel2023gahp}, fuzzy numbers \citep{Karczmarek2021fahp}
or grey numbers \citep{Duleba2022aoga}. The best-known method based
on pairwise comparisons of alternatives is AHP. It is also the most
criticized, although the criticism is often constructive and leads
to many improvements and extensions of the original proposal. Of the
more critical areas of improvement, it is worth mentioning methods
for calculating the vector of weights \citep{Hefnawy2014rodm} including
calculating rankings for incomplete matrices \citep{Bozoki2010ooco,Bozoki2019tlls,kulakowski2020otgm},
methods for measuring inconsistencies \citep{Brunelli2018aaoi,Kulakowski2020iifi}
and others. An interesting critical analysis of the AHP method can
be found in Munier and Hontoria \citep{Munier2021ualo}.

In the PC method, experts, also known as decision-makers, compare
alternatives in pairs. Let $A=\{a_{1},\ldots,a_{n}\}$ be finite set
of alternatives and $E=\{e_{1},\ldots,e_{k}\}$ be a set of experts.
Each expert compares the alternatives with each other to form a set
of comparisons represented in the form of a square matrix $C_{q}=\{c_{ijq}\in\mathbb{R}_{+}:i,j=1,\ldots,n\}$
where $1\leq q\leq k$, and $q$ indicates the number of expert. A
single $c_{ijq}$ denotes the relative importance of the $i$-th alternative
compared to the j-th alternative according to the experts opinion
$e_{q}$. For the sake of legibility of notation, whenever possible
we will omit the expert index $q$ by writing the PC matrix as $C=[c_{ij}]$.
PC matrix is used to synthesize a vector of weights. Let us define
the function resulting from this calculation. 
\begin{defn}
Let $A$ be a set of alternatives. The priority function for $A$
is the mapping $w:A\rightarrow\mathbb{R}_{+}$ assigning a real and
positive number to each alternative.
\end{defn}

The priority vector for $A$ resulting from $C_{q}$ takes the form:
\begin{equation}
w_{q}=\left[w_{q}(a_{1}),\ldots,w_{q}(a_{n})\right]^{T}.\label{eq:weight-vector}
\end{equation}
Many methods have been described in the literature to calculate the
value of priority vector $w$. The first and so far quite popular
is the eigenvalue method (EVM). Saaty proposed this approach in his
seminal paper \citep{Saaty1977asmf}, which consists of calculating
a principal eigenvector and then normalizing it so the sum of entries
is one. Another method is based on calculating the geometric means\footnote{Referred to as geometric mean method (GMM).}
of the matrix rows and taking the result of the $i$-th row as the
priority of the $i$-th alternative \citep{Crawford1987tgmp}. The
result, thus, obtained is also subject to normalization at the end
of the calculation. In addition to the two mentioned, several other
methods for calculating the priority vector exist. For example, Srdjevic
et al. compare the performance of the six popular priority deriving
methods \citep{Srdjevic2023pita}. Another examples of prioritization
methods can be found in \citep{Choo2004acff,Mazurek2022otdo}.

The matrices used in the pairwise comparison method have several specific
properties. Below, we will recall the Frobenius-Perron theorem as
it applies to matrices used in opinion aggregation.
\begin{defn}
\label{def:stochastic-matrix-def} A square matrix $A$ is stochastic
if all of its entries are nonnegative, and the entries of each column
sum to $1$. A matrix is positive if all of its entries are positive
numbers \citep[p. 349]{Margalit2024ila}.
\end{defn}

\begin{thm}[Perron-Frobenus Theorem]
\label{thm:frob-perron} Let $A$ be a positive stochastic matrix.
Then $A$ admits a unique normalized steady state vector $x$, which
spans the $1$-eigenspace. Moreover, for any vector $v_{0}$ with
entries summing to some number $c$, the iterates
\[
v_{1}=Av_{0},v_{2}=Av_{1},\ldots,v_{t}=Av_{t-1},\ldots
\]

approach $c\cdot x$ as $t$ gets large \citep[p. 351]{Margalit2024ila}. 
\end{thm}

\begin{claim}
\label{claim:markov-matrix-props}From the above assertion the following
statements are true \citep[p. 351]{Margalit2024ila}:
\end{claim}

\begin{enumerate}
\item The $1$-eigenspace of a positive stochastic matrix is a line.
\item The $1$-eigenspace contains a vector with positive entries.
\item All vectors approach the $1$-eigenspace upon repeated multiplication
by $A$.
\end{enumerate}

\subsection{Group decision-making}

\subsubsection{Aggregation procedures}\label{subsec:Aggregation-procedures}

When more than one expert participates in decision-making, their opinions
must be aggregated. As with a single PC matrix, the final result should
be a vector of weights assigning real numbers to each alternative.
The aggregation of expert opinions can be done both at the level of
individual comparisons and at the level of individual weight vectors.
An example of the first approach can be the AIJ (aggregating of individual
judgments) \citep{Forman1998aija} procedure, according to which each
comparison of the $i$-th and $j$-th alternatives by each of k experts
is averaged, and then a vector of weights is calculated for the matrix
composed of such averaged values. In addition to AIJ, several optimization
methods directly use expert comparison results. The previously mentioned
model based on minimization of group Euclidean distance (GED) values
\citep{Blagojevic2016haoi} can be an example. A different approach
is the AIP procedure, in which priority vectors for individual experts
are calculated first, and only then are the results aggregated. Compared
to AIJ, the AIP approach has several significant advantages. One undoubted
advantage of the AIP approach is that both geometric and arithmetic
averages can be used to aggregate preferences \citep{Forman1998aija}.
Moreover, the second does not impose the need for a group of experts
to act as a unit, leaving them more freedom to make individual judgments
\citep{Forman1998aija}. It can be easily implemented in the case
of incomplete data from different experts. In such a case, it is enough
to use appropriate prioritization methods for incomplete PC matrices
\citep{Bozoki2010ooco,kulakowski2020otgm} to calculate individual
weight vectors. The weight vectors obtained in this way may then be
subjected to aggregation. In addition, AIP satisfies the Pareto principle
with an arithmetic or a geometric mean. 

Aggregation procedures allow each expert's opinion to be assigned
different strengths (priorities). Thus, experts considered less competent
can have less influence on the final opinion, while more recognized
experts can significantly impact the final outcome. In the following
two sections, we will look at two variants of the AIP method that
allow for assigning different priorities to individual experts.

\subsubsection{Additive aggregation of individual priorities}

Let the vector of weights calculated from the pairwise comparison
matrix $C_{q}$ provided by the $q$-th expert be denoted as follows:

\begin{equation}
w_{q}=\left(\begin{array}{c}
w_{q}(a_{1})\\
w_{q}(a_{2})\\
\vdots\\
w_{q}(a_{n})
\end{array}\right),\,\,\,\text{for}\,\,q=1,\ldots,k.\label{eq:priority-vector-of-q-th-expert}
\end{equation}
Then, let $W$ be a matrix of weights vectors so that 
\[
W=\left(w_{1},\ldots,w_{k}\right),
\]

i.e.

\begin{equation}
W=\left(\begin{array}{ccccc}
w_{1}(a_{1}) & w_{2}(a_{1}) & \cdots & \cdots & w_{k}(a_{1})\\
w_{1}(a_{2}) & w_{2}(a_{2}) & \cdots & \cdots & w_{k}(a_{2})\\
w_{1}(a_{3}) & \vdots & \ddots & \vdots & \vdots\\
\vdots & \vdots & \vdots & \ddots & \vdots\\
w_{1}(a_{n}) & w_{2}(a_{n}) & \cdots & \cdots & w_{k}(a_{n})
\end{array}\right),\label{eq:weights-vectors-matrix-w}
\end{equation}
and, let $p$ be the vector of priorities for individual experts i.e.

\begin{equation}
p=\left(\begin{array}{c}
p_{1}\\
p_{2}\\
\vdots\\
p_{k}
\end{array}\right),\label{eq:expert-prioirty-vector}
\end{equation}
so that $\sum_{i=1}^{k}p_{i}=1$. Then 
\begin{equation}
\text{AAIP}(W,p)\overset{\textit{df}}{=}\left(\begin{array}{c}
\sum_{i=1}^{k}p_{i}w_{i}(a_{1})\\
\sum_{i=1}^{k}p_{i}w_{i}(a_{2})\\
\vdots\\
\vdots\\
\sum_{i=1}^{k}p_{i}w_{i}(a_{n})
\end{array}\right)\label{eq:aaip-def}
\end{equation}
denotes AIP procedure using a weighted arithmetic mean. To simplify
the notation, we will omit the vector $p$ when the priorities of
the experts are equal, i.e. 
\[
\text{AAIP}(W)\overset{\textit{df}}{=}\text{AAIP}\left(W,\left(\frac{1}{n},\ldots,\frac{1}{n}\right)^{T}\right).
\]

\subsubsection{Multiplicative aggregation of individual priorities\protect}\label{subsec:Multiplicative-profile-aggregati}

Let priority vector provided by $q$-th expert be $w_{q}$ (\ref{eq:priority-vector-of-q-th-expert}),
the matrix of priority vectors be given as $W$ (\ref{eq:weights-vectors-matrix-w})
and the vector of expert priorities be $p$ (\ref{eq:expert-prioirty-vector})
so that $\sum_{i=1}^{k}p_{i}=1$. Then
\begin{equation}
\text{GAIP}(W,p)\overset{\textit{df}}{=}\left(\begin{array}{c}
\prod_{i=1}^{k}w_{i}^{p_{i}}(a_{1})\\
\prod_{i=1}^{k}w_{i}^{p_{i}}(a_{2})\\
\vdots\\
\vdots\\
\prod_{i=1}^{k}w_{i}^{p_{i}}(a_{n})
\end{array}\right)\label{eq:gaip-def}
\end{equation}

denotes AIP procedure using a weighted geometric mean. As before,
we will drop the vector $p$ in a situation of equal priorities ie.:
\[
\text{GAIP}(W)\overset{\textit{df}}{=}\text{GAIP}\left(W,\left(\frac{1}{n},\ldots,\frac{1}{n}\right)^{T}\right).
\]

\section{Peer ranking aggregation}

\subsection{Problem description}

The immediate inspiration to take up the topic of aggregation of opinions
in a group of peers was the introduction of changes in the regulations
for rewarding employees of a particular university, according to which
the reward depends on the level of contribution of each author of
a paper. Of course, this level must be numerical (percentage) and
allow for an unambiguous distribution of money between the parties
involved. It must also be based on some compromise because all authors
must sign a declaration agreeing to the assigned share. In practice,
most teams deal with this problem by selecting one person who, acting
as \textquotedbl dictator,\textquotedbl{} proposes the distribution
of \textquotedbl percentages,\textquotedbl{} i.e., funds for the
prize. Depending on the custom, team members may (or may not) negotiate
the proposed share. In the case of teams working on a scientific article,
the role of a dictator is played by the correspondent author, first
author, or the head of the research group.

From the view of the university's financial management, the advantage
of the model with \textquotedbl one person in charge\textquotedbl{}
is the relative simplicity of the solution. The responsibility for
adequately allocating funds is transferred to the person designated
for this purpose. The problem, of course, is most often the lack of
a methodology to make an informed decision on the distribution of
the award, which boils down to adopting the simplest decision-making
model, i.e., dictatorship. The apparent disadvantage of the dictatorial
approach is the arbitrariness of the decisions made. The decision-maker
likes some team members more and others less, which can directly translate
into evaluating their work. He also often overestimates his participation
in the project. On the other hand, the decision-maker is usually the
person most involved in the project and, what goes hand in hand, the
best informed as to the participation and nature of others' contributions.
In the proposed solution, we want to avoid, on the one hand, having
to submit to a dictatorial decision and, on the other hand, to grant
adequate decision-making rights to those who have significantly contributed
to the project's success. 

\subsection{Priority-driven aggregation method}

The above observations led us to propose a peer review model in a
group of peers based on pairwise comparison of alternatives. Group
members have a dual role. They are simultaneously the object of comparison,
i.e. the alternatives $A=\{a_{1},\ldots,a_{n}\}$ in the PC method,
and the experts providing their judgments i.e. $E=\{e_{1},\ldots,e_{n}\}$.
Each of the experts, i.e. team members, provides comparisons in the
form of a PC matrix $C=[c_{ij}]$, where $c_{ij}\in\mathbb{R}_{+}$
in which he assesses the relative level of contribution of the other
members of the group. Based on the collected ratings $C_{1}$,...,$C_{n}$,
vectors of weights $w_{1},\ldots,w_{n}$ for individual experts are
created. The calculated vectors are then subjected to aggregation,
resulting in the formation of a final priority vector

\[
w=\left(\begin{array}{c}
w(a_{1})\\
\vdots\\
\vdots\\
w(a_{n})
\end{array}\right)
\]
reflecting the share of each evaluated person in the award. However,
in line with the observation that those more involved in a project
are often more competent to assess the engagement of others, the priorities
given to expert opinions during aggregation may not be identical.
They must reflect this regularity. Let $p:A\rightarrow\mathbb{R}$
be an expert priority function such that for two experts $e_{i}$
and $e_{j}$ if their final evaluation share meets $w(a_{i})\geq w(a_{j})$
then also their priorities during the aggregation process meet $p(a_{i})\geq p(a_{j})$.
This regularity comes out of the observation that the one who contributed
more to the success than the other has more right than the latter
to decide on the distribution of the reward. In the proposed solution,
we adopted the simplest possible function $p$ i.e. $p(a_{i})=w(a_{i})$.
It grants the right to decide to a team member directly proportional
to his level of involvement. In many cases, it also seems to be the
most equitable. Below, following \citep{Forman1998aija}, we have
proposed two priority-driven aggregation (PDA) methods to support
the proposed solution: multiplicative and additive.

\subsection{Additive priority-driven aggregation method (APDAM)}

When proceeding to aggregate the results, we assume that each team
member $e_{q}\in E$ provided a PC matrix $C_{q}$ based on which
the ranking vector $w_{q}=\left(w_{q}(a_{1}),\ldots,w_{q}(a_{n})\right)^{T}$
was calculated. These vectors form a matrix $W$ (see \ref{eq:weights-vectors-matrix-w}).
Denoting vector of experts' priorities as $p=\left(p(a_{1}),\ldots,p(a_{n})\right)^{T}$
we get an extra condition that the final solution must satisfy:

\begin{equation}
\text{AAIP}(W,\frac{p}{\left\Vert p\right\Vert _{1}})=\left(\begin{array}{c}
p(a_{1})\\
p(a_{2})\\
\vdots\\
\vdots\\
p(a_{n})
\end{array}\right).\label{eq:diad-add-cond}
\end{equation}
where $p(a_{i})>0$ for $i=1,\ldots,n$ and for any $u\in\mathbb{R}^{n}$
holds $\left\Vert u\right\Vert _{1}\overset{\textit{df}}{=}\sum_{i}\left|u_{i}\right|$.
The above condition (\ref{eq:diad-add-cond}) states that the ranking
values of the individual alternatives/experts (i.e., the right side
of the equation) are equal to the priorities (after normalization)
with which the opinions of the individual experts were considered
(the left side of the equation). The equation (\ref{eq:diad-add-cond})
boils down to: 
\begin{equation}
W\frac{p}{\left\Vert p\right\Vert _{1}}=p.\label{eq:diad-eq}
\end{equation}

Since $W$ is a stochastic matrix ie. $W>0$ and $\sum_{j=1}^{n}w_{i}(a_{j})=1$
for $i=1,\ldots,n$, thus, due to Perron-Frobenus theorem (see Theorem
\ref{thm:frob-perron}) there exists a unique and positive eigenvector
$p>0$ with the eigenvalue $1$ whose entries add up to one. Thus
(\ref{eq:diad-eq}) takes the form 
\[
Wp=p.
\]

Moreover the vector $p$ can be calculated iteratively as the limit
of the iteration procedure $W(\ldots W(Wp_{0})\ldots)\rightarrow p$
where entries of the initial vector $p_{0}$ have to sum up to $1$.
It is worth noting that eigenvector calculation algorithms $W$ as
well as global optimization methods can also be used to calculate
the solution. For example condition (\ref{eq:diad-add-cond}) induces
the following optimization problem:

\begin{align}
\min h(W,r)\,\,\,\,\text{s.t.}\nonumber \\
h(W,r)=\sum_{i=1}^{n}\left(\sum_{j=1}^{n}r(a_{j})w_{j}(a_{i})-p(a_{i})\right)^{2}\label{eq:opt-constraint-1}\\
r(a_{i})=\frac{p(a_{i})}{\sum_{j=1}^{n}p(a_{j})}\,\,\,\text{and}\,\,r(a_{i})>0\,\,\text{for}\,\,i=1,\ldots,n.\nonumber 
\end{align}

whose solution is the sought-after vector $p$ (\ref{eq:diad-eq}). 

\subsection{Multiplicative priority-driven aggregation method (MPDAM)\protect}\label{subsec:Multiplicative-profile-aggregation}

In the case of the geometric mean, the condition (\ref{eq:diad-add-cond})
takes the form of 

\begin{equation}
\text{GAIP}(W,\frac{p}{\left\Vert p\right\Vert _{1}})=\left(\begin{array}{c}
p(a_{1})\\
p(a_{2})\\
\vdots\\
\vdots\\
p(a_{n})
\end{array}\right),\label{eq:mpa_condition}
\end{equation}
where $p(a_{i})>0$ for $i=1,\ldots,n$. 

The equivalent non-linear optimization problem is as follows: 

\begin{align}
\min g(W,r)\,\,\,\,\text{s.t.}\nonumber \\
g(W,r)=\sum_{i=1}^{n}\left(\prod_{j=1}^{n}w_{j}^{r(a_{j})}(a_{i})-p(a_{i})\right)^{2}\label{eq:opt-constraint}\\
r(a_{i})=\frac{p(a_{i})}{\sum_{j=1}^{n}p(a_{j})}\,\,\,\text{and}\,\,r(a_{i})>0\,\,\text{for}\,\,i=1,\ldots,n.\nonumber 
\end{align}
The above (\ref{eq:opt-constraint}) can be solved using constrained
global optimization methods. One can also attempt to find the final
vector of weights using a direct iterative algorithm (DIA) bearing
in mind, however, that it may not always converge. The iterative algorithm
works according to the following scheme: 

\begin{algorithm}[h]
\begin{algorithmic}[1]

\alglinenoNew{DIA}

\par \Procedure{DIA}{$W$, $\gamma$, $\epsilon$}

\par \State calculate first approximation of $w'$ using $\text{GAIP}(W)$
\label{dia:init-gaip}

\par \State normalize it so that $\left\Vert w'\right\Vert _{1}=1$
\label{dia:init-normalization}

\par \State  \begin{varwidth}[t]{\dimexpr\linewidth-10ex\relax}
then in at most $\gamma$ iterations repeat the lines (\ref{dia:init-gaip}
- \ref{dia:init-normalization}) verifying each time if a sufficiently
accurate solution has been achieved i.e. $g(W,w')\leq\epsilon$ holds.
If it is so we stop the iteration and return the result $w'$. Otherwise
we continue iterations. If after $\gamma$ iterations the condition
is not satisfied we consider that the algorithm has not found a solution.
\end{varwidth}

\smallskip{}
\par \EndProcedure

\par \alglinenoPush{DIA}\par \end{algorithmic}

\caption*{Direct Iterative Algorithm (DIA)}

\label{alg:dia-1}
\end{algorithm}

If the algorithm DIA returns the correct value of the vector of weights,
this result can be taken as the correct solution. If the calculated
result is too inaccurate we can increase the number of iterations
and try to run the algorithm again. However, if the calculation does
not converge despite increasing the parameters $\gamma$ and $\epsilon$
it is necessary to go for dedicated software using optimization algorithms.
Regardless of the calculation method adopted, we will denote the aggregation
result using this model as $\text{GPDAM}(W)$, where $W$ is a matrix
in which the columns are vectors of weights derived from individual
experts. 

\section{Properties of the methods}

\subsection{Formal properties}\label{subsec:Formal-properties}

When considering various aggregation methods, one of the important
questions is whether they satisfy the Pareto principle \citep{Forman1998aija,Lin2022aotn}
according to which if all experts prefer one alternative over another
then this preference is reflected in the final result. The GAIP and
AAIP aggregation methods satisfy the Pareto postulate \citep{Forman1998aija}.
This property carries over to the MPDAM and APDAM aggregation methods.
Similarly, both methods satisfy the homogeneity condition. 
\begin{prop}
Both modified aggregation methods GAIP and AAIP satisfy the Pareto
principle \citep{Arrow1977scai} i.e. if for any two alternatives
$i,j$ all experts prefer the $i$-th alternative over the $j$-th
alternative i.e. $w_{k}(a_{i})>w_{k}(a_{j})$ for $k=1,\ldots,n$
then this relationship holds in the aggregated vector of weights i.e.
$w(a_{i})>w(a_{j})$.
\end{prop}

\begin{proof}
Let GAIP and AAIP be as defined in (\ref{eq:gaip-def}) and (\ref{eq:aaip-def})
correspondingly. We fix $i,j\in\{1,\ldots,n\}$. For each $p\in\mathbb{R}^{n}$
and for each $k\in\{1,\ldots,n\}$ 
\[
w_{k}(a_{i})>w_{k}(a_{j})\Rightarrow\left(w_{k}^{p_{k}}(a_{i})>w_{k}^{p_{k}}(a_{j})\textnormal{ and }p_{k}w_{k}(a_{i})>p_{k}w_{k}(a_{j})\right),
\]
which implies that 
\[
w_{1}^{p_{1}}(a_{i})\cdot w_{2}^{p_{2}}(a_{i})\cdots w_{n}^{p_{n}}(a_{i})>w_{1}^{p_{1}}(a_{j})\cdot w_{2}^{p_{2}}(a_{j})\cdots w_{n}^{p_{n}}(a_{j})
\]
and 
\[
p_{1}w_{1}(a_{i})+p_{2}w_{2}(a_{i})+\cdots+p_{n}w_{n}(a_{i})>p_{1}w_{1}(a_{j})+p_{2}w_{2}(a_{j})+\cdots+p_{n}w_{n}(a_{j}).
\]
The above inequalities complete the proof. 
\end{proof}
Above that, it is possible to estimate the norm value of both resultant
vectors. 
\begin{prop}
Maps GAIP and AAIP are bounded by $\sqrt{n}$, which is also a Lipschitz
constant for $\text{AAIP}(W,\cdot)$, for any initial matrix $W$.
\end{prop}

\begin{proof}
Using the Cauchy-Schwarz inequality and the inequality between the
geometric and arithmetic means we obtain 
\begin{eqnarray*}
 &  & ||\text{GAIP}(W,p)||=\\
 &  & =\sqrt{(w_{1}^{p_{1}}(a_{1})\cdots w_{n}^{p_{n}}(a_{1}))^{2}+\ldots+(w_{1}^{p_{1}}(a_{n})\cdots w_{n}^{p_{n}}(a_{n}))^{2}}\leq\\
 &  & \leq\sqrt{(p_{1}w_{1}(a_{1})+\ldots+p_{n}w_{n}(a_{1}))^{2}+\ldots+(p_{1}w_{1}(a_{n})+\ldots+p_{n}w_{n}(a_{n}))^{2}}=\\
 &  & =||\text{AAIP}(W,p)||=\\
 &  & =\sqrt{\langle p,w_{1}\rangle^{2}+\ldots+\langle p,w_{n}\rangle^{2}}\leq\sqrt{||p||^{2}||w_{1}||^{2}+\ldots+||p||^{2}||w_{n}||^{2}}=\\
 &  & =||p||\sqrt{||w_{1}||^{2}+\ldots+||w_{n}||^{2}}\leq||p||\sqrt{n}\leq\sqrt{n}.
\end{eqnarray*}

Obviously, the map $\text{AAIP}(W,\cdot)$ is linear, so for each
$p,\hat{p}\in\mathbb{R}^{n}$ 
\[
||\text{AAIP}(W,p)-\text{AAIP}(W,\hat{p})||=||\text{AAIP}(W,p-\hat{p})||\leq\sqrt{n}||p-\hat{p}||.
\]
\end{proof}
Notice that for $p=(\frac{1}{n},\ldots,\frac{1}{n})^{T}$ the norm
\[
||p||=\frac{1}{\sqrt{n}},
\]
so 
\[
||\text{GAIP}(W,p)||\leq||\text{AAIP}(W,p)||\leq1.
\]
On the other hand, if for each $i\in\{1,\ldots,n\}$ we have 
\[
w_{1}(a_{i})=w_{2}(a_{i})=\ldots=w_{n}(a_{i}),
\]
then for each $j\in\{1,\ldots,n\}$ and each $p$ it follows that
\[
\text{GAIP}(W,p)=\text{AAIP}(W,p)=(w_{j}(a_{1}),\ldots,w_{j}(a_{n}))^{T}.
\]

This means that if all the experts scored each other the same way,
then any of the two algorithm will not influence the resulting ranking.

One of the arguments for using the geometric mean to aggregate results,
in addition to satisfying the Pareto principle, is that it satisfies
the homogeneity postulate. According to it \citep{Forman1998aija,Aczel1983pfsr}
if all experts judge a pair of alternatives $t$ times as large as
another pair of alternatives, then the resulting vector also preserve
this ratio. For this reason Aczel and Saaty \citep{Aczel1983pfsr}
argue that for AIJ (see Sec. \ref{subsec:Aggregation-procedures})
the geometric mean must be used. As the weighted geometric and arithmetic
mean also support homogeneity postulate \citep{Aczel1983pfsr} then
also for MPDAM and APDAM this condition holds.
\begin{defn}
The homogeneity condition \citep{Aczel1983pfsr} states:
\[
f(sx_{1},sx_{2},\ldots,sx_{n})=sf(x_{1},x_{2},\ldots,x_{n}),
\]
and $s>0$ and $x_{i},sx_{i}\in R$, where $R$ is a positive set
of numbers containing also their reciprocity values i.e. if $x\in R$
then also $1/x\in R$.
\end{defn}

The homogeneity condition means that the aggregating transformation
f is invariant due to a change in scale or unit (e.g., converting
kilograms to pounds). 
\begin{prop}
MPDAM and APDAM aggregation methods satisfies the homogeneity postulate.
\end{prop}

\begin{proof}
For the additive case we have 
\[
f(w(a_{1}),\ldots,w(a_{n}))\overset{\textit{df}}{=}\sum_{i=1}^{k}\frac{p_{i}}{\sum_{j}p_{j}}w_{i}(a_{k}).
\]

As $\sum_{j}p_{j}=1$ then $f(w(a_{1}),\ldots,w(a_{n}))=\sum_{i=1}^{k}p_{i}w_{i}(a_{k})$.
Hence, for $k=1,\ldots,n$ there is 

\[
f(sw(a_{1}),\ldots,sw(a_{n}))=\sum_{i=1}^{k}sp_{i}w_{i}(a_{k})=\ldots
\]

\[
\ldots=s\sum_{i=1}^{k}p_{i}w_{i}(a_{k})=sf(w(a_{1}),\ldots,w(a_{n}))
\]

Similarly for the multiplicative aggregation 
\[
f(w(a_{1}),\ldots,w(a_{n}))\overset{\textit{df}}{=}\prod_{i=1}^{k}w_{i}^{\frac{p_{i}}{\sum p_{j}}}(a_{k}).
\]

Thus, for $k=1,\ldots,n$ 
\[
f(sw_{1}(a_{k}),\ldots,sw_{n}(a_{k}))=\prod_{i=1}^{k}sw_{i}^{\frac{p_{i}}{\sum p_{j}}}(a_{k})=\ldots
\]

\[
\ldots=\prod_{i=1}^{k}s^{\frac{p_{i}}{\sum p_{j}}}w_{i}^{\frac{p_{i}}{\sum p_{j}}}(a_{k})=\prod_{i=1}^{k}s^{\frac{p_{i}}{\sum p_{j}}}\prod_{i=1}^{k}w_{i}^{\frac{p_{i}}{\sum p_{j}}}(a_{k})=\ldots
\]

\[
\ldots=\left(s^{\frac{p_{1}}{\sum p_{j}}+\ldots+\frac{p_{n}}{\sum p_{j}}}\right)\prod_{i=1}^{k}w_{i}^{\frac{p_{i}}{\sum p_{j}}}(a_{k})=s\prod_{i=1}^{k}w_{i}^{\frac{p_{i}}{\sum p_{j}}}(a_{k})=sf(w(a_{1}),\ldots,w(a_{n})).
\]

\end{proof}
In particular, the proof of homogeneity of MPDAM and APDAM shows that
if the all experts evaluate the ratio between the chosen two alternatives
$a_{x}$ and $a_{y}$ in the same way i.e.
\[
\frac{w_{i}(a_{s})}{w_{i}(a_{t})}=\alpha_{st},\,\,\text{for}\,\,i=1,\ldots,n.
\]
then the same ratio will also be preserved in the aggregate result,
i.e. 

\[
\frac{p(a_{s})}{p(a_{t})}=\alpha_{st}.
\]

In particular, it follows from the above that if all experts agree
on the quantitative relationship between two alternatives then this
relationship holds in the resulting vector. This observation also
means that the Pareto condition is satisfied.

\subsection{Numerical examples}

Discussing the new method of calculating rankings, it is interesting
to observe its behavior on specific simple examples. Below are a few
that will show how the proposed methods work with some interesting
examples. 
\begin{example}
Following the observation at the end of Section \ref{subsec:Formal-properties},
let us consider a situation in which each of the five experts expressed
exactly the same opinions which translated into five identical vectors
of weights. Thus, the considered matrix can look as follows: 
\begingroup\renewcommand*{\arraystretch}{1.2}

\[
W=\left(\begin{array}{ccccc}
\frac{1}{3} & \frac{1}{3} & \frac{1}{3} & \frac{1}{3} & \frac{1}{3}\\
\frac{4}{15} & \frac{4}{15} & \frac{4}{15} & \frac{4}{15} & \frac{4}{15}\\
\frac{1}{5} & \frac{1}{5} & \frac{1}{5} & \frac{1}{5} & \frac{1}{5}\\
\frac{2}{15} & \frac{2}{15} & \frac{2}{15} & \frac{2}{15} & \frac{2}{15}\\
\frac{1}{15} & \frac{1}{15} & \frac{1}{15} & \frac{1}{15} & \frac{1}{15}
\end{array}\right).
\]

\endgroup  
\end{example}

Indeed, all the experts expressed exactly the same opinion. It is
easy to check that the result of aggregation of vectors in $W$ using
MPDAM and APDAM are identical to each of the columns of the $W$:
\[
\text{APDAM}(W)=\text{MPDAM}(W)=\left(\frac{1}{3},\frac{4}{15},\frac{1}{5},\frac{2}{15},\frac{1}{15}\right)^{T}.
\]
\begin{example}
\label{exa:EX5}It is easy to imagine a situation in which two participants
will support each other. Let's consider a situation in which an expert
$e_{1}$ will rate $e_{5}$ highly and vice versa $e_{5}$ will rate
$e_{1}$ highly. The other experts will rate everyone the same. The
matrix $W$ corresponding to such a case will look as follows: 
\begingroup\renewcommand*{\arraystretch}{1.2}
\[
W=\left(\begin{array}{ccccc}
\frac{1}{13} & 1 & 1 & 1 & \frac{9}{13}\\
\frac{1}{13} & 1 & 1 & 1 & \frac{1}{13}\\
\frac{1}{13} & 1 & 1 & 1 & \frac{1}{13}\\
\frac{1}{13} & 1 & 1 & 1 & \frac{1}{13}\\
\frac{9}{13} & 1 & 1 & 1 & \frac{1}{13}
\end{array}\right).
\]
\endgroup  Using the arithmetic mean with equal experts' priorities
to aggregate the results (\ref{eq:gaip-def}) leads to a weight vector
of the form: 
\[
\text{AAIP}(W)=\left(0.222,0.186,0.186,0.186,0.222\right)^{T},
\]
while APDAM gives: 
\[
\text{APDAM}(W)=\left(0.226,0.183,0.183,0.183,0.226\right)^{T}.
\]

Both above ways of aggregation lead to higher expert priorities of
$e_{1}$ and $e_{5}$ than the others, however, in APDAM the priorities
are a bit higher. Indeed, someone considered these experts more important
which further strengthened their opinion. A similar only slightly
stronger effect can be observed with the geometric approach. I.e.,
\[
\text{GAIP}(W)=\left(0.254,0.164,0.164,0.164,0.254\right)^{T},
\]
\[
\text{MPDAM}(W)=\left(0.275,0.15,0.15,0.15,0.275\right)^{T}.
\]

The stronger influence of the opinions of the first and last experts
translates into a larger difference between the weight vectors in
the multiplicative case. Thus, while euclidean distance for additive
method is very small $\left\Vert \text{AAIP}(W)-\text{APDAM}(W)\right\Vert =0.008$,
for the multiplicative one it is an order of magnitude larger: $\left\Vert \text{GAIP}(W)-\text{MPDAM}(W)\right\Vert =0.037$. 
\end{example}

\begin{example}
In example \ref{exa:EX5}, experts $e_{1}$ and $e_{5}$ mutually
assigned the same weight to each other. Thus, in a situation where
the indications will form a cycle, i.e. expert $e_{1}$ will prefer
$e_{2}$ over all others, $e_{2}$ will similarly indicate $e_{3}$,
etc. and the last expert again $e_{1}$, a tie can be expected. Indeed
for 
\begingroup\renewcommand*{\arraystretch}{1.2}
\[
W=\left(\begin{array}{ccccc}
\frac{1}{13} & \frac{9}{13} & \frac{1}{13} & \frac{1}{13} & \frac{1}{13}\\
\frac{1}{13} & \frac{1}{13} & \frac{9}{13} & \frac{1}{13} & \frac{1}{13}\\
\frac{1}{13} & \frac{1}{13} & \frac{1}{13} & \frac{9}{13} & \frac{1}{13}\\
\frac{1}{13} & \frac{1}{13} & \frac{1}{13} & \frac{1}{13} & \frac{9}{13}\\
\frac{9}{13} & \frac{1}{13} & \frac{1}{13} & \frac{1}{13} & \frac{1}{13}
\end{array}\right)
\]

\endgroup leads to the result:
\[
\text{AAIP}(W)=\text{APDAM}(W)=\text{GAIP}(W)=\text{MPDAM}(W)=\left(0.2,\ldots,0.2\right)^{T}.
\]
\end{example}

\begin{example}
One may be concerned about entrusting too much influence to one expert,
who will end up dominating the ranking. Let's try to consider an example
where four experts back-to-back will decisively identify a fifth person
as the ranking winner. The fifth expert, having a separate opinion,
will not give himself (i.e., the fifth alternative) as much weight.
Let's consider an example matrix $W$ corresponding to such a situation:
\begingroup\renewcommand*{\arraystretch}{1.2}
\[
W=\left(\begin{array}{ccccc}
\frac{1}{10} & \frac{1}{15} & \frac{1}{20} & \frac{1}{20} & \frac{1}{10}\\
\frac{1}{20} & \frac{1}{10} & \frac{1}{20} & \frac{1}{15} & \frac{2}{3}\\
\frac{1}{10} & \frac{1}{10} & \frac{1}{5} & \frac{1}{10} & \frac{1}{20}\\
\frac{1}{20} & \frac{1}{15} & \frac{1}{20} & \frac{1}{5} & \frac{1}{20}\\
\frac{7}{10} & \frac{2}{3} & \frac{13}{20} & \frac{7}{12} & \frac{2}{15}
\end{array}\right).
\]
Aggregation using the simple arithmetic mean leads to the weight vector
:
\[
\text{AAIP}(W)=\left(0.073,0.186,0.11,0.083,0.546\right)^{T},
\]
while the APDAM calculation leads to the result
\[
\text{APDAM}(W)=\left(0.081,0.334,0.087,0.065,0.432\right)^{T}.
\]
\endgroup  We can note that the fifth expert did not manage to \textquotedblleft avoid\textquotedblright{}
the fate of the ranking winner. The weight assigned to it by the rest
of the team is still the highest, i.e. $0.432$, although less than
with the standard aggregation method (value: $0.546$). This reduction
occurred at his own request. The strong voice of the fifth expert
had a noticeable effect on increasing the weight of the second expert.
However, it is worth noting that this expert, and with the standard
way of counting, has a significant advantage over the others (without
expert no. $5$). The influence of the last expert is noticeable,
but his influence is effectively limited by the will of the majority. 

In the multiplicative case we get 
\[
\text{GAIP}(W)=\left(0.0858,0.125,0.123,0.0857,0.58\right)^{T},
\]
and 
\[
\text{MPDAM}(W)=\left(0.107,0.261,0.108,0.082,0.44\right)^{T}.
\]
As before, the weight of the fifth expert decreases and the value
of the second expert increases. The weights of the remaining experts
behave similarly. Interestingly, the overall change in the vector
of weights in the second case is slightly larger, i.e. $\left\Vert \text{GAIP}(W)-\text{MPDAM}(W)\right\Vert =0.196$,
where $\left\Vert \cdot\right\Vert $ is an Euclidean distance, while
for the additive case we have $\left\Vert \text{AAIP}(W)-\text{APDAM}(W)\right\Vert =0.189$. 
\end{example}

\section{Montecarlo experiments}

\subsection{Test preparation}

From the perspective of practical use of the proposed methods, the
important thing is how easily one can calculate the desired result.
Calculating the solution using the additive APDAM method is straightforward
and has no difficulties. However, the multiplicative variant of PDAM
requires some computation effort. The presented DIA procedure (Sec.
\ref{subsec:Multiplicative-profile-aggregation}) is fast and straightforward
to implement but does not guarantee success. That is when global optimization
methods can come to the rescue. Their effectiveness, however, is paid
for with longer calculation times. For this reason, we propose a hybrid
approach according to which we first try to use DIA, and in case of
failure, we explore optimization methods (one or more in turn). Following
the hybrid approach idea, in the tests, we execute DIA (where the
maximum number of iterations $\gamma$ is no more than $10^{4}$ and
success is a situation in which $\epsilon<0.1^{4}$), and in case
of failure, we are supported by the library algorithms: simulated-annealing,
Nelder-Mead, and differential evolution techniques. Implementations
of these three algorithms come from the standard library of Wolfram
Mathematica software\footnote{\url{https://reference.wolfram.com/language/tutorial/ConstrainedOptimizationGlobalNumerical.html}}. 

To test the effectiveness of MPDAM, we generated $1,000,000$ matrices
$W$ (\ref{eq:weights-vectors-matrix-w}) in which the columns correspond
to the vectors of weights of individual experts. For the case of two
experts, we generated $10^{5}$ matrices of $2$ by $2$, for three
experts $10^{5}$ matrices of $3$ by $3$, similarly for four experts,
and so on up to the case of $10$ experts, i.e., $10^{5}$ matrices
of $10$ by $10$. We assumed that the values in the vectors obtained
from the experts are normalized, i.e., the sum of the weights in each
matrix column is $1$. 

\subsection{Numerical results}

\begin{figure}
\begin{centering}
\includegraphics[width=0.8\columnwidth]{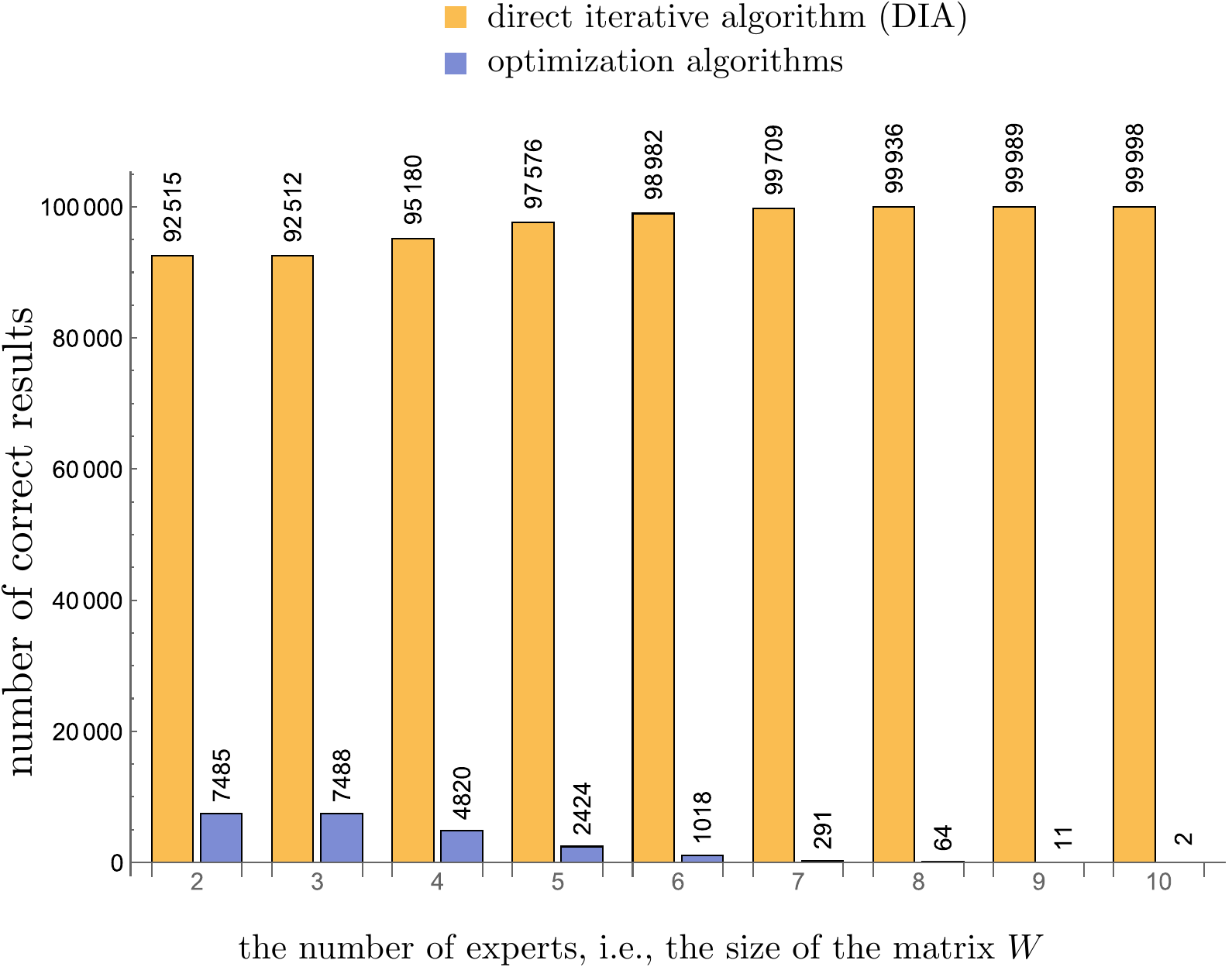}
\par\end{centering}
\caption{The increase of DIA effectiveness along the increase of the number
of experts.}

\label{fig:g-alg-effect}
\end{figure}

In the multiplicative method, DIA does not always find a solution.
The mere existence of a correct solution does not guarantee the convergence
of an iterative algorithm. However, as the number of experts increases,
i.e. the size of the matrix $W$ containing their weights, the DIA's
success frequency increases significantly. For two experts, the DIA
procedure returns the correct solution in $92515$ times out of $10^{5}$
cases, for five already $97576$ times out of $10^{5}$ cases, but
for $10$ experts $99998$ times out of $10^{5}$ tested cases. This
means that in the case of ten experts, only in 2 cases, i.e. $0.002\%$
of all decision scenarios, it was necessary to use optimization algorithms
(Fig. \ref{fig:g-alg-effect}). 

\begin{figure}
\begin{centering}
\includegraphics[width=0.8\textwidth]{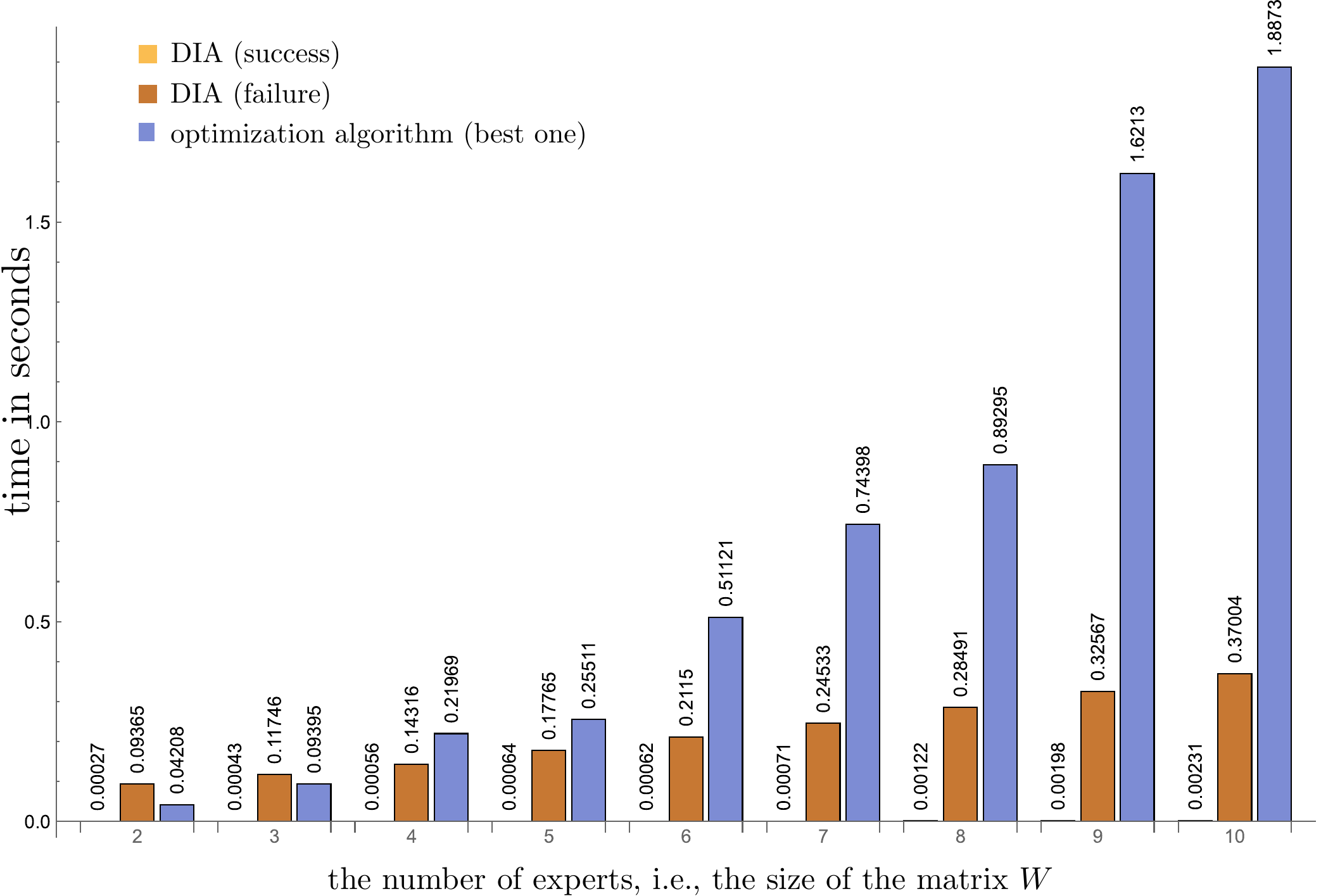}
\par\end{centering}
\caption{Average solution computation time for the DIA algorithm and optimization
algorithms.}

\label{fig:g-alg-time}
\end{figure}

For the computational experiments, a server with an \emph{M2 Max}
processor (\emph{Arm64} architecture) was used. The experimental environment
for the conducted tests was \emph{Wolfram Mathematica$^{\text{TM}}$}
software version $13.3$. Although the tests were conducted in parallel,
the algorithms executed were run sequentially. The DIA procedure used
the following parameter values $\gamma=10000$ and $\epsilon=10^{4}$.
The other algorithms were run with standard parameterization in the
\emph{Wolfram Mathematica} environment, except accuracy goal which
was set to $5.$ This parameter specifies how many effective digits
of accuracy should be sought in the final result. The average times
of sequential execution of the calculations are shown in Figure \ref{fig:g-alg-time}.
As expected, execution times increase as the number of experts increases.
For most cases, the DIA procedure was able to return the correct solution
very quickly. In the case that after \emph{10,000} iterations ($\gamma=10^{4}$)
it was unable to give an acceptable solution i.e. one for which ($\epsilon<0.1^{4}$),
the execution time DIA was significantly higher but still shorter
than the other, global optimization algorithms. The running time of
the professional-grade implementations of optimization algorithms
used for the considered decision models is also increasing along with
the size of the decision model. The average solution running times
for the best optimization algorithm in each case are given in Figure
\ref{fig:g-alg-time}. For larger models (number of experts greater
than $3$), optimization calculations take longer for the given case
than calculations using DIA in case of failure. However, what is essential
from a practical point of view is that as the size of the model increases,
the efficiency of DIA increases rapidly. So, while for a model with
two experts ($W$matrix $2$ by $2$), DIA was unable to return the
correct result in $7\%$ of cases (Fig. \ref{fig:g-alg-effect}),
for a $W$ of size $6$ by $6$, the percentage is just a bit over
$1\%$, and for a $W$ of size $10$ by $10$, it drops to an imperceptible
$0.002\%$. 

Adopting a hybrid strategy according to which we first try to calculate
the solution using DIA and if the algorithm fails then we use optimization
methods allows us to calculate the expected computation time for a
given size of the decision model. Let $\text{ET}(n)$ denotes the
expected time required to compute the solution for the $i$-th model
of size $n$, i.e: 
\[
\text{ET}(n)=\sum_{i\in I_{\textit{DS}}}10^{-5}\cdot\text{DIA}(i,n)+\sum_{i\in I_{\textit{DU}}}10^{-5}\left(\text{DIA}(i,n)+\textit{AVGOPT}(i,n)\right),
\]

where $I_{\textit{DS}}$ is the set of indices of models for which
DIA is able to calculate the solution, $I_{\textit{DU}}$ is the set
of indices of models for which DIA fails (i.e. it performs the maximum
set number of iterations but does not achieve satisfactory accuracy),
$\text{DIA}(i,n)$ is the running time of DIA for the $i$-th model
of size $n$, and $\textit{AVGOPT}(i,n)$ is the average running time
of the optimization algorithms (Nelder-Mead, simulated annealing and
differential evolution) for the $i$-th model of size $n$. 

\begin{figure}
\begin{centering}
\includegraphics[width=0.8\textwidth]{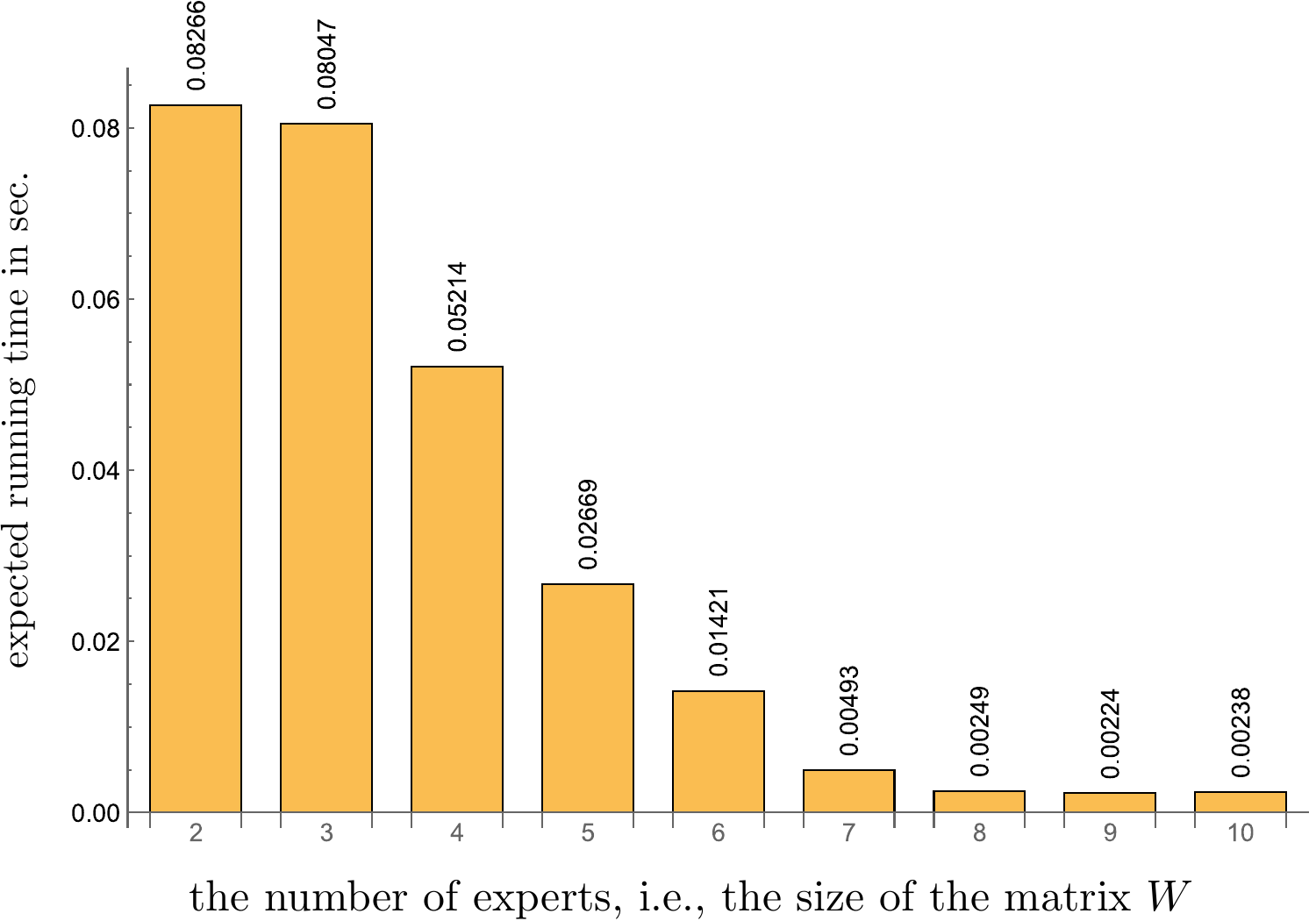}
\par\end{centering}
\caption{The average expected running time using the hybrid strategy combining
the DIA and optimization algorithms }

\label{fig:avg-running-time-hb-strategy}
\end{figure}

It turns out that the average expected running time for hybrid strategy
$\text{ET}(n)$ decreases along the model's size (Fig. \ref{fig:avg-running-time-hb-strategy}).
The decline slows down around the model size $9$ caused by a significant
reduction in the share of global optimization algorithms and begins
a slow, natural grow based on the increasing computation time of the
DIA algorithm. 

\section{Discussion}

The immediate inspiration for the work on the APDAM and GPDAM methods
came from a situation in which one of the authors, playing the role
of \textquotedbl dictator\textquotedbl{} out of necessity, had to
negotiate with the team participants about the division of shares
in a joint project. There was a difference of opinion between the
participants, and each person had arguments to support their position.
However, those more involved in the project had more detailed knowledge
of the nature and scope of the work performed by the team than those
involved less. For this, it seemed natural to favor the opinions of
these very participants more than those whose involvement was small
and fragmented. 

The proposed model attempts to find a compromise by coupling the strength
of a given participant's opinion with the size of his share in the
project. Thus, the opinion of those more involved gains more weight
than those whose involvement is small or negligible. Such situations
can happen in a group of scientists working on a joint paper and in
a team of software engineers creating an IT solution together. In
all these cases, we assume that no external person can objectively
evaluate individuals' commitment. Hence, the need for evaluation is
incumbent on team members. 

The paper presents the opinion aggregation problem in the context
of the pairwise comparison method. For this, the starting point for
considering the aggregation of results was the AIJ and AIP methods
often used in the context of AHP. However, it is worth noting that
using arithmetic and geometric mean for preference aggregation is
not limited to group decision-making in the PC method \citep{Mohd2017amig,Calvo2002aont,Yager1988oowa}.
Therefore, a promising topic for future research seems to be to test
the possibility of solving the posed problem in the context of other
group decision-making methods. 

\section{Summary}

In the work presented here, we proposed two new priority aggregation
methods for group decision-making using the pairwise comparison method,
in which experts' priorities are related to the aggregation result.
For the additive model, we demonstrated the existence of an admissible
solution along with a simple method for calculating the result. For
the multiplicative method, we proposed a method for calculating the
result and performed Montecarlo tests indicating the existence of
a solution. Both aggregation methods satisfy the Pareto principle
and homogeneity condition. The proposed methods correspond to the
situation in which the experts being evaluated are themselves the
object of evaluation. A study example is the problem of determining
the reward distribution for team achievements in a team of peers.
Such an achievement could be the preparation and publication of a
scientific article or the achievement of a milestone in a team of
software engineers. However, the presented approach can have many
other applications, such as distributing bonuses among team members
to achieve organizational or production goals.  In future research,
we will focus on extending the applicability of the proposed model
to other decision-making methods and the properties of the proposed
solutions. In particular, we intend to investigate the safety and
robustness of such models in relation to manipulative behavior.

\section*{Acknowledgment}

The research has been supported by the National Science Centre, Poland,
as a part of the project SODA no. 2021/41/B/HS4/03475. Jacek Szybowski
has also been partially supported by the Polish Ministry of Science
and Higher Education within the internal task of AGH University of
Krakow no. 11.11.420.004. This research was funded in whole or in
part by National Science Centre, Poland 2021/41/B/HS4/03475. For the
purpose of Open Access, the author has applied a CC-BY public copyright
license to any Author Accepted Manuscript (AAM) version arising from
this submission. Special thanks to Prof. Ryszard Szwarc for his comments
on the existence of a solution in the additive case. \bibliographystyle{elsarticle-harv}
\addcontentsline{toc}{section}{\refname}\bibliography{papers_biblio_reviewed}

\end{document}